\documentclass[a4paper,11pt]{article}
\usepackage[utf8]{inputenc}
\usepackage{color}
\usepackage{amssymb}
\usepackage{amsmath}
\usepackage{comment}
\usepackage{graphicx}
\usepackage{epstopdf}
\usepackage{mathtools}
\usepackage{ragged2e}
\usepackage{hyperref}
\hypersetup{
	colorlinks=true,
	linkcolor=blue,
	filecolor=cyan, 
	urlcolor=blue,
	citecolor=red,
} 
\usepackage[titletoc,toc,title]{appendix}
\numberwithin{equation}{section}
\usepackage{cleveref}
\usepackage[margin=2.5cm]{geometry}
\usepackage{cite}
\usepackage{epsfig}
\usepackage{float}
\usepackage{cancel}
\usepackage{braket}
\usepackage{physics}
\usepackage{amsfonts}
\usepackage{enumitem}
\usepackage[font={footnotesize,it}]{caption}
\usepackage{authblk}

\begin{document}
\begin{titlepage}
\title{Covariant holographic reflected entropy in $AdS_3/CFT_2$}

\date{}

\author{Mir Afrasiar\thanks{\noindent E-mail:~ \href{mailto:afrasiar@iitk.ac.in}{afrasiar@iitk.ac.in}}}

\author{Himanshu Chourasiya\thanks{\noindent E-mail:~ \href{mailto:chim@iitk.ac.in}{chim@iitk.ac.in}}}

\author{Vinayak Raj\thanks{\noindent E-mail:~ \href{mailto:vraj@iitk.ac.in}{vraj@iitk.ac.in}}}

\author{Gautam Sengupta\thanks{\noindent E-mail:~ \href{mailto:sengupta@iitk.ac.in}{sengupta@iitk.ac.in}}}

\affil{
Department of Physics\\
			
Indian Institute of Technology\\ 
			
Kanpur, 208016\\ 
			
India }

\maketitle
\begin{abstract}
\noindent
\justify

We substantiate a covariant proposal for the holographic reflected entropy in $CFT$s dual to non-static $AdS$ geometries from the bulk extremal entanglement wedge cross section in the literature with explicit computations
in the $AdS_3/CFT_2$ scenario. In this context we obtain the 
reflected entropy for zero and finite temperature time dependent bipartite mixed states in $CFT_{1+1}$s with a conserved charge dual to bulk rotating extremal and non-extremal BTZ black holes through a replica technique. Our results match exactly with the corresponding extremal entanglement wedge cross section for these bulk geometries in the literature. This constitutes a significant consistency check for the proposal and its possible extension to the corresponding higher dimensional $AdS/CFT$ scenario.

\end{abstract}
\end{titlepage}
\tableofcontents
\pagebreak


\section{Introduction} 
\label{sec1}
\justify

In recent times quantum entanglement has emerged as a significant feature which has received intense research focus in diverse areas from many body systems to black holes and constitutes a rapidly developing field with promises of significant insights. In quantum information theory the entanglement entropy (EE) serves as an appropriate measure for the characterization of  entanglement in bipartite pure states and is defined as the von Neumann entropy of the reduced density matrix although this is usually complicated for quantum many body systems. Interestingly the  EE could be exactly computed in $(1+1)$ dimensional conformal field theories ($CFT_{1+1}$) employing a replica technique as described in \cite{Calabrese:2004eu, Calabrese:2009qy, Calabrese:2009ez}. Subsequently an elegant holographic characterization for the EE was provided by Ryu and Takayanagi \cite{Ryu:2006ef, Ryu:2006bv} involving the area of co dimension two bulk static minimal surfaces in the context of the $AdS/CFT$ correspondence. Their proposal correctly  reproduced the field theoretic replica technique results  of  \cite{Calabrese:2004eu, Calabrese:2009qy, Calabrese:2009ez} for the $AdS_3/CFT_2$ scenario in the appropriate large central charge limit. The above holographic construction was further generalized for non-static bulk $AdS$ geometries in another significant work by Hubeny, Rangamani and Takayanagi (HRT) \cite{Hubeny:2007xt} involving co-dimension two \textit{extremal} surfaces to obtain the covariant EE for time dependent bipartite states in corresponding dual $CFT_d$s. These holographic proposals could be reproduced from the bulk gravitational path integral in a series of subsequent works described in \cite{Fursaev:2006ih, Headrick:2010zt, Casini:2011kv, Faulkner:2013yia, Lewkowycz:2013nqa, Dong:2016hjy}.

However in quantum information theory the EE does not appropriately characterize the entanglement for bipartite mixed states due to the contributions from spurious classical and quantum correlations. Several computable mixed state correlation and entanglement measures such as the entanglement negativity \cite{Vidal:2002zz, Plenio:2005cwa}, odd entanglement entropy \cite{Tamaoka:2018ned}, entanglement of purification \cite{Takayanagi:2017knl, Horodecki:EoP} and balanced partial entanglement \cite{Wen:2021qgx, Camargo:2022mme, Basu:2022nyl, Wen:2022jxr} have been proposed in the existing literature\footnote{There are other mixed state entanglement measures in the quantum information theory literature not mentioned here.}.  An interesting correlation measure termed the \textit{reflected entropy} for mixed state entanglement was introduced in \cite{Dutta:2019gen} which was defined as the entanglement entropy of certain reduced density matrices constructed from the canonical purification of the mixed state in question. Significantly the authors established a novel replica technique to compute the reflected entropy for various bipartite mixed states in $CFT_{1+1}$s. In \cite{Hayden:2021gno}, the authors provided a lower bound to the reflected entropy in terms of the mutual information for the mixed state under consideration.  Furthermore, following the gravitational path integral techniques developed in \cite{Lewkowycz:2013nqa}, the authors in \cite{Dutta:2019gen} also established a duality between the reflected entropy in holographic $CFT$s and the minimal entanglement wedge cross section (EWCS) for the corresponding bulk static $AdS$ geometries. This holographic duality was further extended in \cite{Akers:2021pvd} to the framework of random tensor networks and subsequently was also established for $(1+1)$-dimensional non-relativistic Galilean conformal field theories in \cite{Basu:2021awn,Basak:2022cjs}. 

The developments described above brought into sharp focus the crucial issue of a covariant construction for the reflected entropy of bipartite mixed state configurations in holographic $CFT$s for time dependent scenarios and the generalization of the duality with the corresponding EWCS for the dual bulk non-static $AdS$ geometries. Such a covariant proposal for the reflected entropy was first suggested in the context of the island formulation for black holes in JT gravity using the maximin construction in \cite{Chandrasekaran:2020qtn}. In this work we further address this significant issue and investigate the covariant holographic duality for the reflected entropy of time dependent bipartite mixed states in $CFT$s in terms of the extremal EWCS for the corresponding bulk non static $AdS$ geometries. 
In particular we compute the reflected entropy of various bipartite mixed states in $CFT_{1+1}$s with a conserved charge dual to bulk non extremal and extremal rotating BTZ black holes through an extension to time dependent  scenarios of the replica technique described in \cite{Dutta:2019gen} which verify the holographic duality mentioned above.
Significantly we are able to demonstrate that the replica technique results involving the above generalization matches exactly with twice the extremal EWCS for the bulk rotating BTZ black holes described in the literature. This constitutes a strong substantiation for the identification of the covariant holographic reflected entropy with the extremal EWCS for non static bulk geometries in the $AdS_3/CFT_2$ scenario which should possibly also extend to generic higher dimensional $AdS_{d+1}/CFT_d$ scenarios. However this would require an explicit proof from a covariant bulk gravitational path integral.

This article is organized as follows. In \cref{sec2}, we briefly review the time independent reflected entropy in $CFT_{1+1}$s and its duality with the bulk EWCS. Subsequently in \cref{sec3} we investigate a covariant generalization of the above duality for time dependent mixed states in  $CFT_{1+1}$s with a conserved charge dual to non static bulk rotating BTZ black holes. Next in \cref{sec4} we obtain the covariant reflected entropy for the mixed state of two disjoint intervals in $CFT_{1+1}$s with a conserved charge and compare our results with the bulk extremal EWCS. Following this in \cref{sec5} and \cref{sec6}, we compute the covariant reflected entropy for two adjacent intervals and a single interval respectively. Finally, in \cref{sec7} we present a summary of our work and our conclusions.


\section{Review of reflected entropy}\label{sec2}
In this section we briefly review the essential features of the reflected entropy and its characterization in $CFT_{1+1}$ along with the duality in terms of the static bulk minimal EWCS in the context of the $AdS/CFT$ correspondence as described in \cite{Dutta:2019gen}. To this end we consider a bipartite quantum system $A \cup B$ in a mixed state $\rho_{AB}$. The canonical purification of this state involves the doubling of its Hilbert space to define a pure state $\ket{\sqrt{\rho_{AB}}}_{A B A^* B^*}$ such that
\begin{equation} 
	\rho_{AB} = \text{Tr}_{A^*B^*} \ket{\sqrt{\rho_{AB}}} \bra{\sqrt{\rho_{AB}}}.	
\end{equation}
The reflected entropy $S_{R}(A:B)$ for the bipartite mixed state is defined as the von Neumann entropy of the reduced density matrix $\rho_{AA^{*}}$ as follows 
\begin{equation}\label{def.}
S_{R}(A:B) = S_{vN} (\rho_{AA^{*}})_{\sqrt{\rho_{AB}}} \,\,\,,
\end{equation}
where the reduced density matrix $\rho_{AA^{*}}$ is given as
\begin{equation}
\rho_{AA^{*}}= \text{Tr}_{BB^*}\ket{\sqrt{\rho_{AB}}} \bra{\sqrt{\rho_{AB}}}.
\end{equation}

Interestingly, the authors in \cite{Dutta:2019gen} developed a novel replica technique involving two replica indices $(m,n)$ which could be utilized to compute the reflected entropy for a bipartite mixed state described by subsystems $A\equiv [z_1,z_2]$ and $B\equiv[z_3,z_4]$ in a $CFT_{1+1}$. The R\'enyi reflected entropy was defined in terms of the partition functions $Z_{n,m}$ on an $nm$-sheeted replica manifold and $Z_{1,m}$ on an $m$-sheeted replica manifold as \cite{Dutta:2019gen}
\begin{equation}\label{defination}
S_{n}(AA^{*})_{\psi_{m}}=\frac{1}{1-n}\log \frac{Z_{n,m}}{(Z_{1,m})^n}=\frac{1}{1-n} \log \frac{\left<\sigma_{g_{A}}(z_{1})\sigma_{g_{A}^{-1}}(z_{2})\sigma_{g_{B}}(z_{3})\sigma_{g_{B}^{-1}}(z_{4})\right>_{CFT^{\otimes mn}}}{\left(\left<\sigma_{g_{m}}(z_{1})\sigma_{g_{m}^{-1}}(z_{2})\sigma_{g_{m}}(z_{3})\sigma_{g_{m}^{-1}}(z_{4})\right>_{CFT^{\otimes m}}\right)^n} \, .
\end{equation}
The state $ \ket{\psi_{m}}$ above is defined on an $m$-replication of the original complex plane with $m \in 2\mathbb{Z^{+}}$ and $\sigma_{g}$'s are twist fields defined on the replicated manifold. The scaling dimensions of the operators $\sigma_{g^{}_A}, \sigma_{g^{}_B}$ and $\sigma_{g_{m}}$ are given as \cite{Dutta:2019gen}
\begin{equation}
h\equiv h_A =h_B =\frac{n c}{24} \left(m-\frac{1}{m}\right)=\bar{h}_A=\bar{h}_B, \quad \quad\quad\quad\, h_{m}= \frac{c}{24}\left(m-\frac{1}{m}\right).
\end{equation} 
The reflected entropy for such bipartite states may finally be obtained in the replica limit $n \to 1$ and $m \to 1$ as
\begin{equation}\label{SR-def}
	S_{R}(A:B)=\lim_{n,m \to1} S_{n}(AA^{*})_{\psi_{m}}.
\end{equation}
We should note here that the order of the limits in the above expression is a subtle issue which has been investigated in \cite{Akers:2021pvd, Akers:2022max, Kusuki:2019evw} where it was shown that the order of the analytic continuation of the replica indices $n \to 1$ and $m \to 1$ is inequivalent in the large central charge limit. It was suggested that first $n \to 1$ limit should be implemented and subsequently $m \to 1$. The alternate ordering of the limits could result in an incorrect selection of the dominant channel in the large $c$ limit which could further violate certain quantum information bounds on the reflected entropy. In this article, we will be using the appropriate order $n \to 1$ and subsequemtly $m \to 1$ to obtain the reflected entropy as suggested in \cite{Akers:2021pvd, Akers:2022max, Kusuki:2019evw}.

Significantly it was shown in \cite{Dutta:2019gen} that the reflected entropy for the bipartite state configuration of disjoint intervals in a $CFT_{1+1}$ on a fixed time slice matched exactly in the large central charge limit, with twice the minimal EWCS for the dual bulk static $AdS_3$ geometry in the framework of the $AdS_3/CFT_2$ correspondence. The proposed holographic duality between the reflected entropy and the minimal EWCS was then established in \cite{Dutta:2019gen, Akers:2022max} through the gravitational path integral technique developed in \cite{Lewkowycz:2013nqa}, in the context of the general $AdS/CFT$ correspondence for static bulk geometries. This holographic duality could be expressed as follows 
\begin{equation}\label{duality}
S_R(A:B) = 2 E_W(A:B),
\end{equation}
where the $E_W(A:B)$ is the bulk minimal EWCS corresponding to the bipartite state $\rho_{AB}$.

\section{Covariant reflected entropy in $CFT_{1+1}$ with a conserved charge} \label{sec3}

\subsection{Covariant reflected entropy}

In this subsection we define the reflected entropy for time dependent mixed state configurations in a $CFT_{1+1}$. For this purpose we consider the time dependent bipartite mixed state of two boosted disjoint intervals $A \equiv [(x_1,t_1),(x_2,t_2)]$ and $B \equiv [(x_3,t_3),(x_4,t_4)]$ in a $CFT_{1+1}$ on the complex plane and the covariant reflected entropy for this configuration may be defined as follows 
\begin{equation}\label{covref. defination}
S_{R}(A:B)=\lim_{n,m \to1} \frac{1}{1-n} \log \frac{\left<\sigma_{g_{A}}(x_{1},t_{1})\sigma_{g_{A}^{-1}}(x_{2},t_{2})\sigma_{g_{B}}(x_{3},t_{3})\sigma_{g_{B}^{-1}}(x_{4},t_{4})\right>_{CFT^{\otimes mn}}}{\left(\left<\sigma_{g_{m}}(x_{1},t_{1})\sigma_{g_{m}^{-1}}(x_{2},t_{2})\sigma_{g_{m}}(x_{3},t_{3})\sigma_{g_{m}^{-1}}(x_{4},t_{4})\right>_{CFT^{\otimes m}}\right)^n} \, ,
\end{equation}
where $\sigma_g$'s are time dependent twist operators inserted at the end points of the intervals. From the $AdS_3/CFT_2$ correspondence the $CFT_{1+1}$ in the time dependent bipartite mixed state is dual to a bulk non static $AdS_3$ geometry. 

We now investigate the covariant version of the holographic duality between the reflected entropy for time dependent mixed states in a $CFT_{1+1}$ and the corresponding extremal EWCS inspired by \cite {Hubeny:2007xt, Chandrasekaran:2020qtn} for the dual non static bulk $AdS_3$ geometry given as
\begin{equation}\label{Duality}
S_R(A:B) = 2 E^\text{ext}_{W}(A:B).
\end{equation}
In the following subsections we substantiate the covariant holographic duality described in \cref{Duality} by computing the reflected entropy of time dependent bipartite mixed states in a $CFT_{1+1}$ with a conserved  charge dual to bulk non extremal and extremal rotating BTZ black holes and show that our results exactly match with twice the corresponding extremal EWCS computed in \cite{KumarBasak:2021lwm}. This would constitute a significant consistency check for the covariant generalization of the duality between the holographic reflected entropy and the extremal EWCS. 

\subsection{$CFT_{1+1}$ with a conserved charge}
In this subsection we briefly describe the structure of a $CFT_{1+1}$ with a conserved charge \cite {Caputa:2013eka, Caputa:2013lfa} dual to bulk non extremal and extremal rotating BTZ black holes. The conserved charge of the $CFT_{1+1}$ may be interpreted the angular momentum of the bulk rotating BTZ black hole. The corresponding $CFT_{1+1}$ is then defined on a $(1+1)$ dimensional twisted cylinder\footnote {By twisted cylinder we mean that the compactifications of the holomorphic and anti holomorphic parts are different which lead two different inverse temperatures $\beta_{+}$ and $\beta_{-}$ as shown in \cref{partition function}.} whose Euclidean partition function is described as \cite{Hubeny:2007xt, Banados:1992wn, Banados:1992gq, Caputa:2013lfa,Caputa:2013eka},
\begin{equation}\label{partition function}
\mathbb{Z}=\text{Tr}\left(e^{-\beta(H+\Omega_EJ)}\right)\equiv \text{Tr} \left(e^{- \beta_{+} L_{0} - \beta_{-} \bar{L}_{0}}\right ).
\end{equation}
Here the Hamiltonian is given by $ H = L_{0} + \bar{L}_{0} $ where $ L_{0} $ and $ \bar{L}_{0} $ are the standard Virasoro zero modes, $\beta$ is the inverse temperature, $J$ is the bulk angular momentum which is the conserved charge in the $CFT_{1+1}$ and $\Omega_{E}= -i \Omega$ is the Euclidean angular potential. From the $AdS_3/CFT_2$ correspondence the $CFT_{1+1}s$ with a conserved charge at a finite temperature are dual to bulk non extremal rotating BTZ black holes whose temperature and angular potential may be expressed in terms of the horizon radii $r_{\pm}$  as\cite{Caputa:2013lfa,Banados:1992wn,Banados:1992gq}
\begin{equation}\label{beta-omega}
\frac{1}{\beta}= \frac{r_{+}^{2}-r_{-}^{2}}{2 \pi r_{+}} ~, \quad \Omega= \frac{r_{-}}{r_{+}}.
\end{equation}
The inverse temperature $\beta$ is then expressed for the holomorphic and the anti holomorphic sectors as follows \cite{Caputa:2013eka, Caputa:2013lfa}
\begin{equation}\label{tempsss}
\beta_+ = \beta(1+i\Omega_E)~, \quad \beta_- = \beta(1-i\Omega_E)~.
\end{equation}
In the replica technique it is now required to compute the correlation functions of the twist fields for the $CFT_{1+1}$ on the twisted cylinder in order  to obtain the covariant reflected entropy for bipartite mixed state configuration. To this end it is possible to map the correlation function on the complex plane to the corresponding twisted cylinder through the conformal transformations given by
\begin{equation}\label{twisted-cylinder-transf}
w=\frac{\beta_{+}}{2 \pi}\log{z}, \quad \bar{w}=\frac{\beta_{-}}{2 \pi}\log{\bar{z}},
\end{equation}
where $z$ and $w$ denotes the coordinates on the complex plane and the twisted cylinder respectively.

For $CFT_{1+1}$s with a conserved charge at a zero temperature the corresponding bulk dual geometries are described by extremal rotating BTZ black holes for which the inverse temperature $\beta \to \infty$ and the angular potential $\Omega \to 1$ as observed from \cref{beta-omega}.  In this limit one of the inverse temperatures namely the $\beta_{+}$ goes to infinity and the other one $\beta_{-}$ remains finite. For this case the conformal transformation which maps the correlation functions on the complex plane to the twisted cylinder are given as
\begin{equation}\label{twisted-cylinder-transf(II)}
w=z, \quad \bar{w}=\frac{\beta_{-}}{2 \pi}\log{\bar{z}}.
\end{equation}
Here $\beta_{-}$ may be interpreted as the inverse of an effective Frolov-Thorne (FT) temperature which corresponds to the degeneracy of the ground state of an extremal BTZ black hole \cite{Frolov:1989jh, Caputa:2013lfa}. Under a conformal map $z \to  w$, the four point twist correlator in \cref{covref. defination} transforms as 
\begin{equation}\label{Refl-Disj-Tr.}
\begin{aligned}
&\left<\sigma_{g^{}_A}(w_1,\bar{w}_1)\sigma_{g_A^{-1}}(w_2,\bar{w}_2)\sigma_{g^{}_B}(w_3,\bar{w}_3)\sigma_{g_B^{-1}}(w_4,\bar{w}_4)\right>_{\beta_{\pm}}=\prod_{i=1}^{4} \left(\frac{dw_i}{dz_i}\right)^{-h_{i}}\left(\frac{d\bar{w}_i}{d\bar{z}_i}\right)^{-\bar{h}_{i}}\\
& \quad \hspace{6 cm}
\left<\sigma_{g^{}_A}(z_1,\bar{z}_1)\sigma_{g_A^{-1}}(z_2,\bar{z}_2)\sigma_{g^{}_B}(z_3,\bar{z}_3)\sigma_{g_B^{-1}}(z_4,\bar{z}_4)\right>_{\mathbb{C}}\,,
\end{aligned}
\end{equation}
where $h_{i}$ and $\bar{h}_{i}$ are the scaling dimensions of the twist field located at $z= z_i$. In the subsequent sections we will compute the covariant reflected entropy for the mixed state configurations described by two disjoint intervals, two adjacent intervals and a single interval in the $CFT_{1+1}$ with a conserved charge at both zero and finite temperatures and compare our results in the large central charge limit with the extremal bulk EWCS computed earlier in \cite{KumarBasak:2021lwm} to substantiate our covariant holographic reflected entropy construction.

\section{Covariant reflected entropy for two disjoint intervals} \label{sec4}

\begin{figure}[H]
	\centering
	\includegraphics[scale=.60]{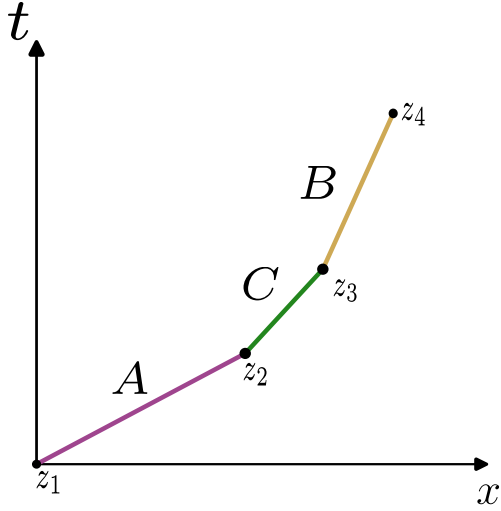}
	\caption{Schematics of two boosted disjoint intervals $A\equiv[z_1,z_2]$ and $B\equiv[z_3,z_4]$ on a complex plane. }
	\label{fig1}
\end{figure}

In this section we will compute the covariant reflected entropy for the time dependent bipartite mixed state configuration of two disjoint intervals in $CFT_{1+1}$s with a conserved charge. In order to observe the effect of the conserved charge on the corresponding twist field correlators it is required to implement the time dependence by considering two boosted disjoint intervals $A \equiv [z_1, z_2]$ and $B \equiv [z_3, z_4]$ on the complex plane where $z_i= (x_i,t_i)$ as shown in \cref{fig1}. The lengths of the two intervals $A$ and $B$ may be expressed in terms of the coordinates $(x_i,t_i)$  as $l_A= \sqrt{\lvert x_2-x_1\rvert^{2}- \lvert t_2-t_1 \rvert^{2}}$ and $l_B= \sqrt{\lvert x_4-x_3\rvert^{2}- \lvert t_4-t_3 \rvert^{2}}$.

In the $t$-channel the numerator of \cref{covref. defination} may be expressed in terms of the conformal blocks as 
\begin{equation}\label{four-point-expansion}
\left<\sigma_{g^{}_A}(z_1,\bar{z}_{1})\sigma_{g_A^{-1}}(z_2,\bar{z}_{2})\sigma_{g^{}_B}(z_3,\bar{z}_{3})\sigma_{g_B^{-1}}(z_4,\bar{z}_{4})\right>_{\mathrm{CFT}^{\bigotimes mn}}=z_{14}^{2h}\bar{z}_{14}^{2{h}}z_{23}^{2h}\bar{z}_{23}^{2{h}}
\sum_p C_p^2 \mathcal{F}_{mn} (\eta) \mathcal{\bar{F}}_{mn}(\bar{\eta}) ,
\end{equation} 
where $\eta= \frac{z_{12}z_{34}}{z_{13}z_{24}}$, $\bar{\eta}=\frac{\bar{z}_{12}\bar{z}_{34}}{\bar{z}_{13}\bar{z}_{24}}$ are the cross ratios. Here $C_p^2$ is the OPE coefficient of the corresponding three point function, computed in \cite{Dutta:2019gen} as  $C_p=(2m)^{-2h/n-2\bar{h}/n}$ and $\mathcal{F}_{mn}$ and $\mathcal{\bar{F}}_{mn}$ are the Virasoro conformal blocks. In general the explicit form of the conformal blocks are unknown. However we may define the semi-classical large central charge limit here as $mnc\to \infty$ for which the parameters $\epsilon\equiv\frac{6h}{mnc}$ and $\epsilon_{p}\equiv\frac{6h_{p}}{mnc}$ stays fixed, where $h_{p}$ is the scaling dimension of the twist operator providing the dominant contribution in the conformal block expansion. In such large central charge limit, the Virasoro conformal blocks $\mathcal{F}_{mn}$  and $\mathcal{\bar{F}}_{mn}$ exponentiate as follows \cite{Zamolodchikov1987}
\begin{equation}\label{log-F}
\begin{aligned}
\log \mathcal{F}_{mn}\left(\eta\right) \approx -\frac{mnc}{6} f \left(\eta\right)   , \\ 
\log \mathcal{\bar{F}}_{mn}\left(\bar{\eta}\right) \approx -\frac{mnc}{6} \bar{f} \left(\bar{\eta}\right).
\end{aligned}
\end{equation}
In the $t$-channel and in the large $c$ limit, the dominant contribution to the four point twist correlator in \cref{four-point-expansion} arises from the conformal block corresponding to the primary twist field $\sigma_{g^{}_B g_A^{-1}}$ with the scaling dimension
\begin{equation}
h_p = h_{B A^{-1}}= \frac{2c}{24} \left(n-\frac{1}{n}\right).
\end{equation}
The perturbative expansion of  $f$ in $\epsilon$ and $\epsilon_{AB} = \frac{6 }{mnc}h_{B A^{-1}}$ may be given as \cite{Fitzpatrick:2014vua}
\begin{equation}\label{f_+}
f\left(\eta\right)=\epsilon_{AB}\log\left(\frac{1+\sqrt{\eta}}{1-\sqrt{\eta}}\right)+\mathrm{higher\, order\, terms}\,.
\end{equation}
We will now utilize the above expression to compute the covariant reflected entropy for finite and zero temperature scenarios. 

\subsection{Finite temperature}
We now consider two boosted disjoint intervals $A$ and $B$ of lengths $l_A$ and $l_B$ in a $CFT_{1+1}$s with a conserved charge at a finite temperature which are dual to bulk non-extremal rotating BTZ black holes. To obtain the reflected entropy for this configuration we need to utilize the conformal map given in \cref{twisted-cylinder-transf} which modifies the cross-ratios as follows
\begin{equation}\label{finite temp. cross ratio}
	\xi_{+}= \frac{\sinh\left(\frac{\pi  w_{12}}{\beta_{+}}\right)\sinh\left(\frac{\pi  w_{34}}{\beta_{+}}\right)}{\sinh\left(\frac{\pi  w_{13}}{\beta_{+}}\right)\sinh\left(\frac{\pi  w_{24}}{\beta_{+}}\right)},
	\quad\quad\quad
	\xi_{-}= \frac{\sinh\left(\frac{\pi  \bar{w}_{12}}{\beta_{-}}\right)\sinh\left(\frac{\pi \bar{w}_{34}}{\beta_{-}}\right)}{\sinh\left(\frac{\pi \bar{w}_{13}}{\beta_{-}}\right)\sinh\left(\frac{\pi \bar{w}_{24}}{\beta_{-}}\right)}.
\end{equation}
Now by using \cref{covref. defination,four-point-expansion,f_+} the reflected entropy in the large central charge limit for this configuration may be computed as follows  
\begin{equation}
S_R{(A:B)}= \frac{c}{6} \log \left[\frac{1+\sqrt{\xi_{+}}}{1-\sqrt{\xi_{+}}}\right]+\frac{c}{6} \log \left[\frac{1+\sqrt{\xi_{-}}}{1-\sqrt{\xi_{-}}}\right].
\end{equation}
We observe that in the proximity limit for the $t$-channel $(\xi_{\pm} \to 1)$, our result is consistent with the covariant holographic duality given in \cref{Duality} in terms of the corresponding extremal EWCS computed for the bulk non extremal rotating BTZ black hole \cite{KumarBasak:2021lwm} in the context of the $AdS_3/CFT_2$ scenario.

\subsection{Zero temperature}
In this subsection we will obtain the reflected entropy for the mixed state configuration of two disjoint intervals in a $CFT_{1+1}$ with a conserved charge at zero temperature. In this case, utilizing \cref{twisted-cylinder-transf(II),Refl-Disj-Tr.,four-point-expansion,f_+} we obtain the time dependent reflected entropy to be
\begin{equation}
S_R{(A:B)}= \frac{c}{6} \log \left[\frac{1+\sqrt{\eta}}{1-\sqrt{\eta}}\right]+ \frac{c}{6} \log \left[\frac{1+\sqrt{\xi_{-}}}{1-\sqrt{\xi_{-}}}\right],
\end{equation}
where $\eta$ and $\xi_{-}$ are the zero and finite temperature cross ratios respectively. The first term in the above expression represents the reflected entropy for two disjoint intervals in a $CFT_{1+1}$ without a conserved charge at zero temperature, while the second term describes the correction in the reflected entropy due to an effective FT temperature. Additionally, we notice that in the proximity limit, the reflected entropy obtained above matches with twice the corresponding extremal EWCS computed in the context of the dual bulk extremal rotating BTZ black hole in \cite{KumarBasak:2021lwm} which is once again consistent with the covariant holographic duality mentioned in \cref{Duality}.

\section{Covariant reflected entropy for two adjacent intervals} \label{sec5}
\begin{figure}[H]
	\centering
	\includegraphics[scale=.60]{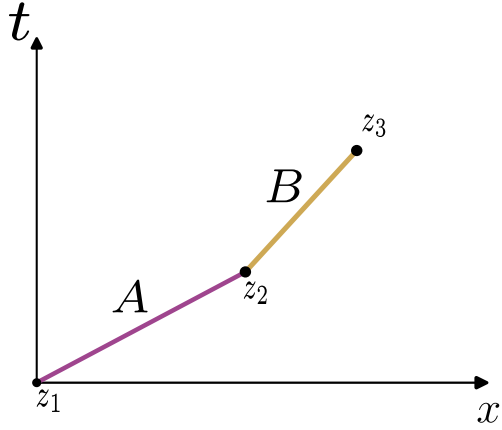}
	\caption{Schematics of two boosted adjacent intervals $A\equiv[z_1,z_2]$ and $B\equiv[z_2,z_3]$  on a complex plane}
	\label{fig2}
\end{figure}

In this section we compute the covariant reflected entropy for the bipartite mixed state of two adjacent intervals in  $CFT_{1+1}$s with a conserved charge. For this purpose we consider two boosted adjacent intervals $A$ and $B$ of lengths $l_A$ and $l_B$ respectively in a $CFT_{1+1}$ with a conserved charge defined on a twisted cylinder. The end point coordinates of the adjacent intervals are given by $A \equiv [z_1, z_2]$ and $B \equiv [z_2, z_3]$ as shown in \cref{fig2}, where $z_i= (x_i,t_i)$ are the coordinates on the complex plane. 
As described earlier the correlation functions of a $CFT_{1+1}$ with a conserved charge defined on a twisted cylinder may be related to a $CFT_{1+1}$ on the complex plane through the conformal transformation described in \cref{twisted-cylinder-transf}. Under this conformal map the corresponding three point twist correlator for the above configuration, required to compute the time dependent reflected entropy, transforms as
\begin{equation}\label{refl-three-point-transf}
\begin{aligned}
\left<\sigma_{g^{}_A}(w_1,\bar{w}_1)\sigma_{g^{}_B g_A^{-1}}(w_2,\bar{w}_2)\sigma_{g_B^{-1}}(w_3,\bar{w}_3)\right>_{\beta_{\pm}}=&\prod_{i=1}^{3} \left(\frac{dw_i}{dz_i}\right)^{-h_{i}}\left(\frac{d\bar{w}_i}{d\bar{z}_i}\right)^{-\bar{h}_{i}}\\
&  \left<\sigma_{g^{}_A}(z_1,\bar{z}_1)\sigma_{g^{}_B g_A^{-1}}(z_2,\bar{z}_2)\sigma_{g_B^{-1}}(z_3,\bar{z}_3)\right>_{\mathbb{C}}.
\end{aligned}
\end{equation}
In the following subsections we obtain the covariant reflected entropy for two adjacent intervals at finite and zero temperatures.

\subsection{Finite temperature}
In this subsection, we discuss the case of the two adjacent intervals in the $CFT_{1+1}$ with a conserved charge at a finite temperature $1/\beta$. The form of the corresponding three point twist correlator on the complex plane is given by
\begin{equation}\label{three pt.function}
\left<\sigma_{g^{}_A}(z_1,\bar{z}_1)\sigma_{g^{}_B g_A^{-1}}(z_2,\bar{z}_2)\sigma_{g_B^{-1}}(z_3,\bar{z}_3)\right>_{\mathbb{C}}= \frac{C_{A,BA^{-1},B^{-1}}^{2}}{z_{12}^{h_{B A^{-1}}} z_{23}^{h_{B A^{-1}}}z_{13}^{2h-h_{B A^{-1}}} \bar{z}_{12}^{h_{B A^{-1}}} \bar{z}_{23}^{h_{B A^{-1}}}\bar{z}_{13}^{2h-h_{B A^{-1}}}},
\end{equation}
where $C_{A,BA^{-1},B^{-1}}$ is the OPE coefficient obtained in \cite{Dutta:2019gen}. The covariant reflected entropy for the two adjacent intervals under consideration may be obtained by utilizing \cref{twisted-cylinder-transf,three pt.function,refl-three-point-transf} in \cref{covref. defination} 
as follows
\begin{equation}
\begin{aligned}	
S_R{(A:B)}=&\frac{c}{6}\log\Bigg[\bigg(\frac{\beta_{+}}{\pi \epsilon}\bigg)\frac{\sinh{\big(\frac{\pi w_{12}}{\beta_{+}}\big)}\sinh{\big(\frac{\pi w_{23}}{\beta_{+}}\big)}}{\sinh({\frac{\pi w_{13}}{\beta_{+}} })}\Bigg]\\
+&\frac{c}{6}\log\Bigg[\bigg(\frac{\beta_{-}}{\pi \epsilon}\bigg)\frac{\sinh{\big(\frac{\pi \bar{w}_{12}}{\beta_{-}}\big)}\sinh{\big(\frac{\pi \bar{w}_{23}}{\beta_{-}}\big)}}{\sinh({\frac{\pi \bar{w}_{13}}{\beta_{-}} })}\Bigg]+\frac{c}{3}\log 4,
\end{aligned}	
\end{equation}
where $\epsilon$ is a UV cut off of the corresponding $CFT_{1+1}$. We again note that in the large central charge limit the reflected entropy for this case also precisely matches with twice the corresponding extremal EWCS obtained in \cite{KumarBasak:2021lwm}. This is once again consistent with the covariant holographic duality between the time dependent  reflected entropy and the bulk extremal EWCS described in \cref{Duality}. 

\subsection{Zero temperature}
In this subsection we study the covariant reflected entropy for the two boosted adjacent intervals in a $CFT_{1+1}$ with a conserved charge at zero temperature. For this case also the $CFT_{1+1}$ is defined on a twisted cylinder and utilizing \cref{twisted-cylinder-transf(II),refl-three-point-transf,covref. defination} the covariant reflected entropy for the mixed state under consideration may be obtained as
\begin{equation}
S_R{(A:B)}=\frac{c}{6}\log\Bigg[\frac{w_{12} w_{23}}{\epsilon w_{13}}\Bigg]
+\frac{c}{6}\log\Bigg[\bigg(\frac{\beta_{-}}{\pi \epsilon}\bigg)\frac{\sinh{\big(\frac{\pi \bar{w}_{12}}{\beta_{-}}\big)}\sinh{\big(\frac{\pi \bar{w}_{23}}{\beta_{-}}\big)}}{\sinh({\frac{\pi \bar{w}_{13}}{\beta_{-}} })}\Bigg]+\frac{c}{3}\log 4.
\end{equation}
Here the first term describes the reflected entropy for the configuration of two adjacent intervals in a $CFT_{1+1}$ without a conserved charge at zero temperature and the second term is the correction to the reflected entropy due to the effective FT temperature ${1}/{\beta_{-}}$ of the extremal rotating BTZ black hole. Once again we observe that in the large central charge limit the above expression explicitly matches with twice the corresponding extremal EWCS computed in \cite{KumarBasak:2021lwm}
which is consistent with the covariant holographic duality given in \cref{Duality}.

\section{Covariant reflected entropy for a single interval}\label{sec6}
In this section we compute the covariant reflected entropy for the mixed state configuration of a single interval $A \equiv [0,l_A]$ in a $CFT_{1+1}$ with a conserved charge\footnote {Note that in $CFT_{1+1}$ with a conserved charge even the single interval at zero temperature is a mixed state due to the effective temperature arising from the vacuum degeneracy.}. It has been shown in \cite{Basu:2022nds} that the presence of an infinite branch cut for the reflected entropy of a single interval in a $CFT_{1+1}$ with anomaly at a finite temperature requires a tripartition with the single interval enclosed between two large but finite auxiliary intervals $B_1 \equiv [-L, 0]$ and $B_2 \equiv [l_A, L]$ and finally a bipartite limit has to implemented through $\pm L\to \pm\infty$. Naturally this is also required to compute the reflected entropy for the single interval in our case of a $CFT_{1+1}$ with a conserved charge as shown in the figure below. 

\begin{figure}[H]
	\centering
	\includegraphics[scale=.40]{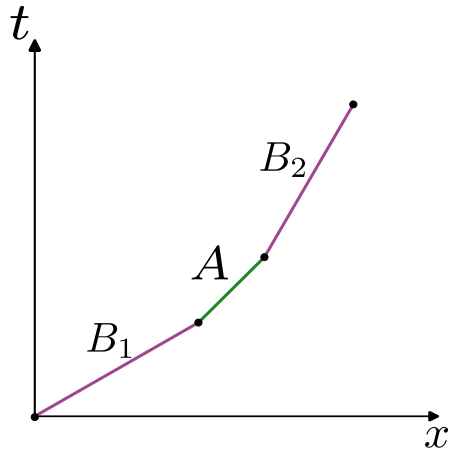}
	\caption{Schematics of a single interval $A$ adjacent to two auxiliary intervals $B_1$ and $B_2$  on a complex plane}
	\label{fig3}
\end{figure}
The covariant reflected entropy for this configuration may then be defined as
\begin{equation}\label{Sn_R-correct-singleT}
S_R\left(A:B\right)=\lim_{L \to \infty}\lim_{n,m \to1}\frac{1}{1-n}\log \frac{\left<\sigma_{g_B}(-L)\sigma_{g^{-1}_B g_A^{}}(0)\sigma_{g^{}_B g_A^{-1}}(l_A)\sigma_{g_B^{-1}}(L)\right>_{\mathrm{CFT}_{\beta_\pm}^{\bigotimes mn}}}{\bigg( \left<\sigma_{g_m^{}}(-L)\sigma_{g_m^{-1}}(L)\right>_{\mathrm{CFT}_{\beta_\pm}^{\bigotimes m}}\bigg)^n},
\end{equation}
where the subscript $\beta_\pm$ denotes that the four point twist correlator is being evaluated on a twisted cylinder instead of the complex plane. Here the bipartite limit $B \equiv B_1 \cup B_2 \to A^c \ (L \to \infty) $ has to be employed after the replica limit $n,m \to 1$ as described in \cite{Basu:2022nds}. We may compute the four point twist correlator in \cref{Sn_R-correct-singleT} from the corresponding four point twist correlator on the complex plane by utilizing the conformal map given in \cref{twisted-cylinder-transf}. The four point twist correlator on the complex plane may then be expressed as \cite{Calabrese:2014yza, Basu:2022nds}
\begin{align} \label{four-point-single}
\Big<\sigma_{g_B}(z_1)\sigma_{g^{-1}_B g_A^{}}(z_2)&\sigma_{g^{}_B g_A^{-1}}(z_3)\sigma_{g_B^{-1}}(z_4)\Big>_{\mathrm{CFT}^{\bigotimes mn}}\notag \\&=k_{mn}\left(\frac{1}{z_{14}^{2h}z_{23}^{2h_{{B} A^{-1}}}}\frac{\mathcal{F}_{mn}(\eta)}{\eta^{h_{{B} A^{-1}}}}\right)\left(\frac{1}{\bar{z}_{14}^{2h}\bar{z}_{23}^{2h_{{B} A^{-1}}}}\frac{\bar{\mathcal{F}}_{mn}(\bar{\eta})}{\bar{\eta}^{h_{{B} A^{-1}}}}\right), 
\end{align}
where $k_{mn}$ is a constant and $\eta, \bar{\eta}$ are the cross ratios. The non universal functions of the cross ratios $\mathcal{F}_{mn}(\eta)$ and $\bar{\mathcal{F}}_{mn}(\bar{\eta})$ in the limits $\eta, \bar{\eta} \to 1$ and $\eta, \bar{\eta} \to 0$ are given as \cite{Basu:2022nds}
\begin{equation}
\mathcal{F}_{mn}(1)=\bar{\mathcal{F}}_{mn}(1)=1, \hspace{5mm} \mathcal{F}_{mn}(0)=\bar{\mathcal{F}}_{mn}(0)=C_{mn},
\end{equation}
where $C_{mn}$ is a non universal constant depending upon the full operator content of the theory. In the next subsections we will utilize the above expressions to compute the covariant reflected entropy for the single interval in the $CFT_{1+1}$ with a conserved charge at finite and zero temperatures.

\subsection{Finite temperature}
In this subsection we obtain the covariant reflected entropy for a single interval in a $CFT_{1+1}$ with a conserved charge at a finite temperature. The four point twist field correlator  required to compute the time dependent reflected entropy on the twisted cylinder, may now be obtained from those on the complex plane through the conformal transformation given in \cref{twisted-cylinder-transf} as follows
\begin{equation}\label{4pt-singleT}
\begin{aligned}
&\left<\sigma_{g_B}(-L)\sigma_{g^{-1}_B g_A^{}}(0)\sigma_{g^{}_B g_A^{-1}}(l_A)\sigma_{g_B^{-1}}(L)\right>_{\mathrm{CFT}_{\beta_\pm}^{\bigotimes mn}} \\
&\qquad\qquad\qquad=k_{mn}\left[ \left(\frac{\beta_+}{\pi\epsilon}\right)\sinh\left(\frac{2\pi L}{\beta_+}\right) \right]^{-2h} \left[ \left(\frac{\beta_+}{\pi\epsilon}\right)\sinh\left(\frac{\pi l_A}{\beta_+}\right) \right]^{-2h_{{B} A^{-1}}}\frac{{\mathcal{F}}_{mn}({\xi_+})}{\xi_{+}^{h_{{B} A^{-1}}}} \\
&
\qquad\qquad\qquad\qquad \times \left[ \left(\frac{\beta_-}{\pi \epsilon}\right)\sinh\left(\frac{2\pi L}{\beta_-}\right) \right]^{-2h} \left[ \left(\frac{\beta_-}{\pi\epsilon}\right)\sinh\left(\frac{\pi l_A}{\beta_-}\right) \right]^{-2h_{{B} A^{-1}}}\frac{\bar{\mathcal{F}}_{mn}({\xi_-}\,)}{\xi_{-}^{\,h_{{B} A^{-1}}}},
\end{aligned}
\end{equation}
where $\xi_{+}$ and $\xi_{-}$ are the finite temperature cross ratios defined in \cref{finite temp. cross ratio}. In the bipartite limit they have the following form 
\begin{equation}\label{modified cross ratios}
\lim_{L\to \infty} \xi_{+}= e^{-\frac{2\pi l_A}{\beta_+}}, \,\,\,
\lim_{L\to \infty} \xi_{-}= e^{-\frac{2\pi l_A}{\beta_-}}.
\end{equation}
Using \cref{4pt-singleT,modified cross ratios} in \cref{Sn_R-correct-singleT} and implementing the bipartite limit $L \to \infty$ subsequent to the replica limit $n,m \to 1$ we may obtain the reflected entropy for the single interval as  
\begin{equation}
S_R{(A:B)}= \frac{c}{3} \log\left[\left(\frac{\beta_{+}\beta_{-}}{\pi^2 \epsilon^2}\right)\sinh\left(\frac{\pi \l_{A}}{\beta_{+}}\right)\sinh\left(\frac{\pi \l_{A}}{\beta_{-}}\right)\right]-\frac{\pi c\, l_A}{3 \beta_{+}}-\frac{\pi c\, l_A}{3 \beta_{-}}+ f\left(e^{\frac{-2 \pi l_{A}}{\beta_{+}}}\right)+\bar{f}\left(e^{\frac{-2 \pi l_{A}}{\beta_{-}}}\right).
\end{equation}
Here the non universal functions $f(\xi_+)$ and $\bar{f}(\xi_{-})$ are given by
\begin{equation}
f(\xi_+)=\lim_{n,m \to 1}\log [{\mathcal{F}}_{mn}({\xi_{+}})], \qquad \qquad
\bar{f}(\xi_-)=\lim_{n,m \to 1}\log [\bar{\mathcal{F}}_{mn}(\xi_{-})]\,.
\end{equation}
It is possible to express the above equation in a more suggestive form as follows 
\begin{equation}
S_{R}(A:B)= 2 \left[S_{A}-S_{A}^\text{th}\right]+f\left(e^{\frac{-2 \pi l_{A}}{\beta_{+}}}\right)+\bar{f}\left(e^{\frac{-2 \pi l_{A}}{\beta_{-}}}\right),
\end{equation}
where $S_{A}$ denotes the covariant entanglement entropy of a single interval $A$ and $S_{A}^\text{th}$ denotes its thermal entropy. Once again we observe that in the large central charge limit the universal part of the above result matches with twice the corresponding extremal EWCS up to an additive constant\footnote{The additive constant may be extracted from the non universal functions $f$ and $\bar{f}$ corresponding to the four point twist correlator in \cref{four-point-single} through a proper large central charge monodromy analysis of a six point twist correlator in appropriate channels as described in \cite{ Malvimat:2017yaj}.\label{additive-constant}} computed in \cite{KumarBasak:2021lwm}. 

\subsection{Zero temperature}
In this subsection we compute the covariant reflected entropy for a single interval in a $CFT_{1+1}$ with a conserved charge at zero temperature. Once more for this case we need to consider the two auxiliary intervals $B_1$ and $B_2$ and compute the four point function in \cref{Sn_R-correct-singleT} on a twisted cylinder. As described earlier we will again employ the conformal transformation in \cref{twisted-cylinder-transf(II)} to obtain the four point twist field correlator on the twisted cylinder as follows
\begin{equation}\label{4pt-singleT(II)}
\begin{aligned}
\Big<\sigma_{g_B}(-L)\sigma_{g^{-1}_B g_A^{}}(0)\sigma_{g^{}_B g_A^{-1}}&(l_A)\sigma_{g_B^{-1}}(L)\Big>_{\mathrm{CFT}_{\beta_\pm}^{\bigotimes mn}} =k_{mn}(2L)^{-2h} (l_{A})^{-2h_{{B} A^{-1}}}\frac{{\mathcal{F}}_{mn}({\eta})}{\eta^{h_{{B} A^{-1}}}} \\
& \times \left[ \left(\frac{\beta_-}{\pi \epsilon}\right)\sinh\left(\frac{2\pi L}{\beta_-}\right) \right]^{-2h} \left[ \left(\frac{\beta_-}{\pi \epsilon}\right)\sinh\left(\frac{\pi l_A}{\beta_-}\right) \right]^{-2h_{{B} A^{-1}}}\frac{\bar{\mathcal{F}}_{mn}({\xi_-}\,)}{\xi_{-}^{\,h_{{B} A^{-1}}}},
\end{aligned}
\end{equation}
where $\eta$ and $\xi_{-}$ are zero temperature and finite temperature cross ratios. Once more using \cref{4pt-singleT(II),modified cross ratios} in \cref{Sn_R-correct-singleT} and implementing the replica and the bipartite limits, the covariant reflected entropy for this case may be obtained to be
\begin{equation}
S_R{(A:B)}= \frac{c}{3}\log(\frac{l_{A}}{\epsilon})  +\frac{c}{3} \log\left[\left(\frac{\beta_{-}}{\pi \epsilon}\right)\sinh\left(\frac{\pi \l_{A}}{\beta_{-}}\right)\right]-\frac{\pi c \,l_A}{3 \beta_{-}}+\bar{f}\left(e^{\frac{-2 \pi l_{A}}{\beta_{-}}}\right).
\end{equation}
As in the previous subsection we may express the above equation as 
\begin{equation}
S_{R}(A:B)= 2 \Big[S_{A}-S_{A}^\text{FT}\Big]+\bar{f}\left(e^{\frac{-2 \pi l_{A}}{\beta_{-}}}\right).
\end{equation}
Here we observe that the covariant reflected entropy for this case involves the elimination of  $S_{A}^\text{FT}= \frac{\pi c l_{A}}{6 \beta_{-}}$ which is an effective thermal entropy of the subsystem $A$ at the FT temperature $1/\beta_{-}$ and corresponds to the entropy of degeneracy of the ground state. Once again we note that the above expression matches with the extremal EWCS obtain in \cite{KumarBasak:2021lwm} apart from an additive constant (c.f. \cref{additive-constant}).

\section{Summary}\label{sec7}
To summarize, in this article we have investigated a covariant generalization of the holographic duality between the reflected entropy of time dependent bipartite mixed states in $CFT_{1+1}$s with the EWCS of  the corresponding bulk dual non static $AdS_3$ geometries. In this context we have computed the reflected entropy for various time dependent bipartite mixed states in $CFT_{1+1}$s with a conserved charge dual to bulk non extremal and extremal rotating BTZ black holes from a generalized replica technique. The bipartite mixed states in the $CFT_{1+1}$ with a conserved charge were described by the boosted configurations of two disjoint, two adjacent and a single interval at finite and zero temperatures. Remarkably the field theory replica technique results for the reflected entropy of these time dependent states matched exactly (modulo an additive constant for the single interval configuration) with the results for the bulk extremal EWCS computed in the literature which constitute a strong consistency check for the covariant holographic duality mentioned earlier.

In this connection we started with the time dependent bipartite mixed state configurations described by two boosted disjoint intervals in $CFT_{1+1}$s with a conserved charge defined on a twisted cylinder at finite and zero temperatures. The corresponding reflected entropy for this mixed state was described by a four point twist field correlator using a replica technique, which could be computed from a monodromy analysis in the large central charge limit. It was observed that in the proximity limit of the two disjoint intervals the reflected entropy in the large central charge limit matched exactly with twice the corresponding extremal EWCS for the dual non extremal and extremal bulk rotating BTZ black holes consistent with the corresponding covariant holographic duality.

Subsequently we extended our computations to mixed state configurations described by two boosted adjacent intervals in  $CFT_{1+1}$s with a conserved charge at finite and zero temperatures. Once again the reflected entropy for the above time dependent mixed state matched exactly with twice the extremal EWCS in the literature for the corresponding bulk dual rotating BTZ black holes further confirming the covariant holographic duality. Interestingly we could demonstrate that our results for the two adjacent intervals could be reproduced from those for the disjoint intervals above, in an appropriate limit constituting a further consistency check.

Finally we have computed the reflected entropy for bipartite mixed state configurations of a boosted single interval in $CFT_{1+1}$s with a conserved charge at finite and zero temperatures. For this case it was required to employ a replica construction from the literature, involving two large but finite auxiliary intervals with the single interval sandwiched between these. Interestingly once again our results modulo an additive constant, confirmed the covariant holographic duality between the reflected entropy and twice the EWCS for the corresponding bulk rotating BTZ black hole geometries. The additive constant may be determined in the large central charge limit from the non universal function associated with the four point twist correlator  through a monodromy analysis for specific channels of a six point twist correlator.

The results obtained in this article constitutes a significant consistency check and a strong substantiation of the covariant extension of the holographic duality between the reflected entropy of time dependent bipartite states in $CFT_{1+1}$s dual to non static bulk $AdS_3$ geometries and the extremal EWCS. It would be extremely interesting to examine this covariant holographic construction in the context of other non static explicitly time dependent bulk $AdS_3$ geometries. It is tempting to suggest that the covariant holographic duality between the reflected entropy and the bulk extremal EWCS extends to generic higher dimensional $AdS_{d+1}/CFT_d$ scenarios also. This would require an explicit proof from a covariant bulk gravitational path integral perspective which is a non trivial and involved issue.  These constitute exciting future open directions in this context.  

\section*{Acknowledgment}
We would like to thank Debarshi Basu for several insightful discussions. The work of GS is partially supported by the Dr. Jagmohan Garg Chair Professor position at the Indian Institute of Technology, Kanpur.

\bibliographystyle{utphys}

\bibliography{ref}

\end{document}